\documentclass[aps,prl,twocolumn]{revtex4}

\usepackage{textcomp}
\usepackage{graphicx}
\usepackage{latexsym}
\usepackage{amsfonts}
\usepackage{amsmath}
\usepackage[T1,T2A]{fontenc}
\usepackage[utf8]{inputenc}

\begin{document}
\title{Energy gap in tunneling spectroscopy: effect of the chemical potential shift}
\author{N.I. Fedotov}
\author{S.V. Zaitsev-Zotov}
\email[]{serzz@cplire.ru}
\affiliation{Kotel'nikov IRE of RAS, 125009, Moscow, Mokhovaya 11, bld.7}

\begin{abstract}
We study the effect of a shift of the chemical potential level on the tunneling conductance spectra. In the systems with gapped energy spectra, significant chemical-potential dependent distortions of the differential tunneling conductance curves, $dI/dV$, arise in the gap region. 
An expression is derived for the correction of the $dI/dV$, which in a number of cases was found to be large. The sign of the correction depends on the chemical potential level position with respect to the gap. The correction of the $dI/dV$ associated with the chemical potential shift has a nearly linear dependence on the tip-sample separation $z$ and vanishes at $z\to 0$.
 \end{abstract}
\maketitle

\section{Introduction}
The scanning tunneling spectroscopy (STS) is widely used to study the energy structure of atomically clean surfaces of solids with spatial resolution down to the atomic level. In many cases the question of the presence of an energy gap in the energy spectrum of the surface states plays a crucial role. For instance, the two-dimensional electron gas on the surfaces of semiconductors \cite{Himpsel,Hasegawa}, monolayers of various layered materials  (as a review see \cite{Liu})
 superconductor surfaces \cite{Zhang} and especially the topological insulators \cite{Molenkamp} 
  are examples of the objects where the STS may help to distinguish between the different possible surface phases.  As the surface properties are usually affected by numerous parameters such as, for example, the synthesis conditions, substrate doping, or presence of adsorbate affecting the doping level of the surface states, the question of the effect of the chemical potential level on the STS results is also of great interest.  

In the scanning tunneling spectroscopy the  differential conductance,   $dI/dV$, of the tunneling junction (sometimes  normalized by $I/V$) is used as a measure for the local density of states (LDOS),   $\rho_s(eV)$ (see, for example, \cite{Feenstra,Bai,Chen,STS,Voigtlaender} and references therein). LDOS is usually considered to be proportional to the differential tunneling conductance, $\rho_s(eV)\propto dI/dV$, and, consequently, nonzero $dI/dV$ indicates nonzero $\rho_s(eV)$ and vice versa. In practice, however, this method leads to contradictory results. In some cases, such as superconductors, the energy gap region does correspond to zero differential tunneling conductance  (up to  thermal smearing). In other cases, \emph{e. g.}, in the  Bi$_2$Se$_3$ topological insulator, the tunneling spectroscopy of atomically clean surfaces  does not yield zero $dI/dV$ at the Dirac point \cite{Urazhdin}, moreover the values at this point, obtained on different samples, differ significantly (see, \emph{e. g.}, \cite{Dai}).
 
 It is however known, that the relation $\rho_s(eV)\propto dI/dV$ holds with the finite precision, and a vast amount of literature is dedicated to   the recovery of $\rho_s(E)$ from $I(V)$ tunneling characteristics  (see \cite{Passoni,Koslowski, Ukraintsev} and references therein). 

For the purposes of the present work it is enough to consider a simple model  \cite{Simmons} of the planar tunneling junction conductance in the trapezoidal barrier approximation. Numerous refinements of this model, taking into account image charges, tunneling probe shape and other factors (see, \emph{e. g.}, \cite{Feenstra,Bai,Chen,STS,Voigtlaender,Passoni}), do not change  the conclusions of the present work essentially.

Tunneling current in the framework of this model  is given by
\begin{equation}
\begin{aligned}
I=A\int_{-\infty}^{\infty}\rho_s(E)\rho_t(E-eV)T(E,V,z)\\
\left(f(E-eV)-f(E)\right)dE,
\end{aligned}
\label{eq:i}
\end{equation}
where $\rho_t(E)$ is the density of states of the probe,  $T(E,V,z)$ is the tunneling barrier transmission, $f(E)$ is the distribution function.

 For the sake of simplicity let  $e \equiv 1$, and let us consider the case of zero temperature. Then the expression for the tunneling current (\ref{eq:i}) becomes
 $$
I=A\int_{0}^{V}\rho_s(E)\rho_t(E-V)T(E,V,z)dE,
$$
 and for the differential conductance we have:
\begin{equation}
\begin{aligned}
\frac{dI}{dV}= A\rho_s(V)\rho_t(0)T(V,V,z) + \\
+A\int_{0}^{V}\rho_s(E)\frac{\partial{}}{\partial{}V}\left[ \rho_t(E-V)T(E,V,z)\right] dE.
\end{aligned}
\label{eq_dva_slagaemyh}
\end{equation}
Here the first term is directly connected with the local density of states, but contains  $T(V,V,z)$ as a factor, which depends on $V$ and strongly depends on $z$. 
In the simplest case the dependence of  $T$ on $V$ is neglected, by assuming $T=e^{-\kappa z}$, and $dI/dV$ is normalized by $T$ or simply by the value of $dI/dV$ at a particular point.
Often to take into account the dependence of $T$ on $V$ the normalization by $I/V$ is used. Such a normalization may lead to divergences at the edges of the energy gap in case of a gapped LDOS, so in such cases $I/V$ is averaged over an interval  $\Delta V$ \cite{Feenstra1}. The symmetrical form of equation (\ref{eq:i}) with respect to the tip and the sample and its normalization $F=A_tT(V,V,z)+A_sT(0,V,z)$ were proposed by Ukraintsev. This method yields the density of  unoccupied states of the sample at the positive  $V$ and the density of unoccupied states of the tip at the negative $V$ \cite{Ukraintsev}.
The second term in (\ref{eq_dva_slagaemyh}) is a contribution, brought about by  the change of the barrier transmission with the applied voltage, as well as by a non-constant density of states of the probe.
In practice this term is usually dropped because it is assumed to be small or to form a smooth background, insignificant in determination of the peak positions in the LDOS. 
 A more accurate determination of  $\rho_s(E),$ taking into account this term,  is the subject of a number of papers, \emph{e. g.}  \cite{Koslowski,Passoni}. 
Koslowski \emph{et. al.}  \cite{Koslowski} using the  WKB approximation for a trapezoidal barrier reduced the problem to a set of integral equations and proposed an iterative procedure to solve it. 
 The WKB parameters estimation and effects of non-constant tip density of states are analyzed in the work of Passoni \emph{et. al.} \cite{Passoni}.

We are interested in the differences in the tunneling conduction spectra (besides a shift in voltage), arising  in otherwise identical systems with different chemical potentials of samples.
It follows from equation (\ref{eq_dva_slagaemyh}) that if the chemical potential position does not fall outside the energy gap, 
 both the current, and its correction are equal to 0 and the presence of a gap in the density of states is correctly determined by the STS. 
This fact explains, for example, a successful use of the tunneling spectroscopy in the superconductor studies, where the gap is opened on the Fermi level.  
As we  show below, a completely different situation may arise in the study of semiconducting surfaces, 
where the chemical potential can have different values and be outside the energy gap.   

In the present work we analyze the effect of the chemical potential position of the sample on the differential tunneling conductance. We show, that the effect of the chemical potential shift does not come down to just a voltage shift of the  $\frac{dI}{dV}(V)$ curves.
 We derive an expression for the correction of the tunneling junction conductance, which in a number of cases was found to be significant
and can have both positive (at $V<0$) and negative (at $V>0$) values. 
A method of identification and mitigation of this effect on the basis of the analysis of the tip-sample separation dependence is proposed.

\section{Effect of the chemical potential shift}

Let us consider conditions under which such a nonzero correction emerges and how it affects STS results obtained at different chemical potential values. For this purpose  we calculate the change in the differential tunneling conductance  if the chemical potential is shifted by $\delta\mu$.  

\begin{figure}
\includegraphics[width=0.22\textwidth]{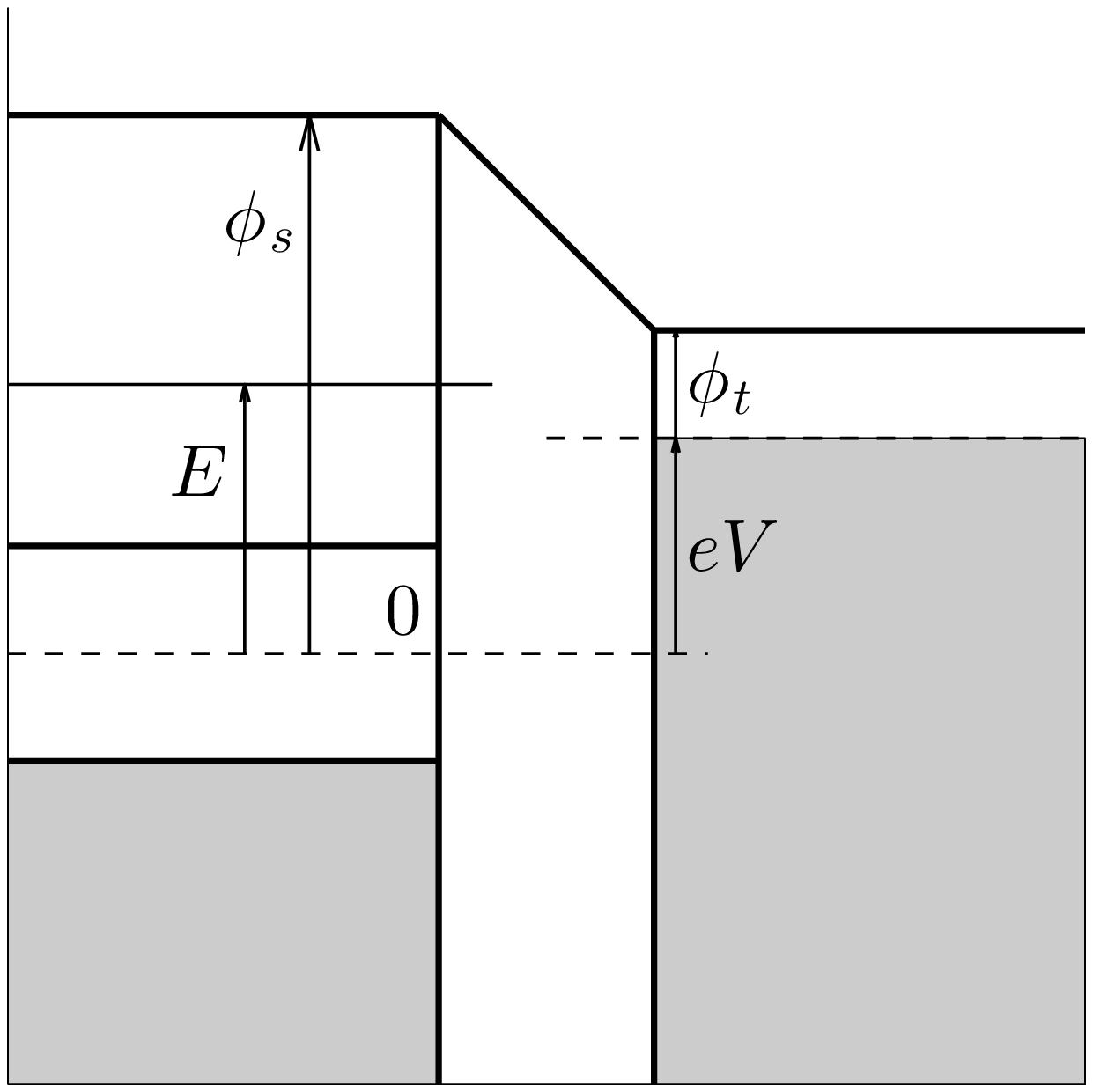}
\includegraphics[width=0.22\textwidth]{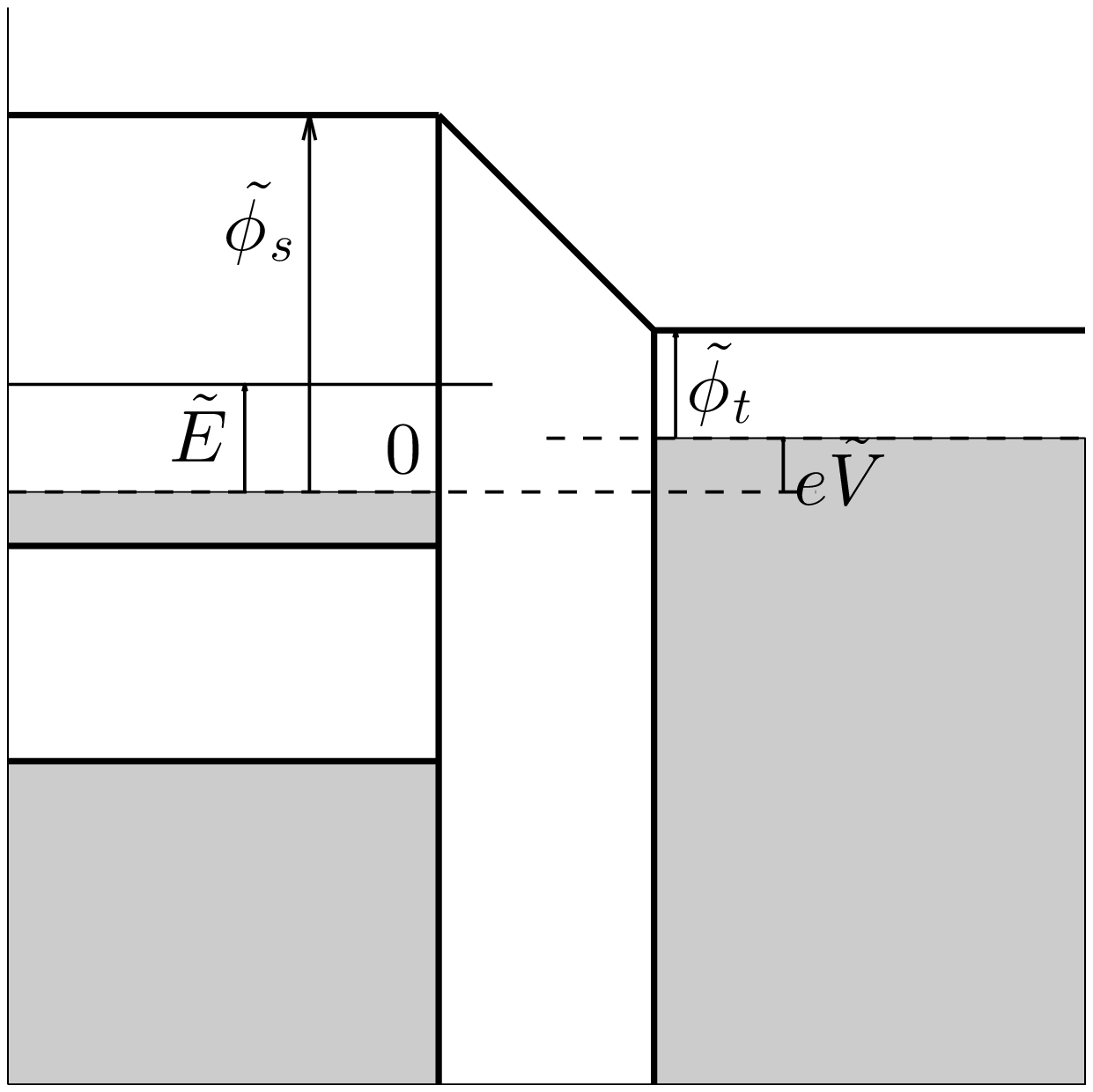}
\caption{Energy diagrams of two tunnel junctions with different chemical potential of the sample.}
\label{model}
\end{figure}

Consider two identical systems, different in the value of the chemical potential only (Fig. \ref{model}). 
The parameters of these systems are related as follows :
$$
\begin{aligned}
\tilde{E}=E-\delta\mu,\\
\tilde{V}=V-\delta\mu,\\
\tilde{\phi_s}=\phi_s-\delta\mu,\\
\tilde{\phi_t}=\phi_t,\\
\tilde{\rho_s}(\tilde{E})=\rho_s(E),\\
\tilde{\rho_t}(\tilde{E}-\tilde{V})=\rho_t(E-V),\\
\tilde{T}(\tilde{E},\tilde{V})=T(E,V),\\
\end{aligned}
$$
where $\phi_t$ and $\phi_s$ are the work functions of the tip and the sample correspondingly. Tunneling current in the second system can be expressed through the tunneling current in the first one:
$$
\tilde{I}=I-A\int_{0}^{\delta\mu}\rho_s(E)\rho_t(E-V)T(E,V,z)dE.
$$
Note, that the difference between the currents is associated only with states with energies between the chemical potential levels of the two systems.
After differentiating we obtain
$$
\begin{aligned}
\frac{d\tilde{I}}{d\tilde{V}}=\frac{dI}{dV}+A\int_{0}^{\delta\mu}\rho_s(E)\frac{d\rho_t(E-V)}{dV}T(E,V,z)dE-\\
-A\int_{0}^{\delta\mu}\rho_s(E)\rho_t(E-V)\frac{\partial T}{\partial V}(E,V,z)dE
\end{aligned}
$$
It is evident that the chemical potential shift results not only in a voltage shift of the differential conductance curve,  $G(V)$, by $\delta \mu/e$, but also in the appearance of an additional contribution, $\Delta G$, changing the value of $dI/dV$.
This contribution contains a part, associated with a non-constant tip density of states, and a part, caused by the change of the tunneling barrier transmission with  applied voltage. Let us focus on the latter by assuming
$\rho_t\equiv 1$. Then
\begin{equation}
\Delta G=-A\int_{0}^{\delta\mu}\rho_s(E)\frac{\partial T}{\partial V}(E,V,z)dE.
\label{eq:deltaG}
\end{equation}
Note that since  ${\partial T(E,V,z)}/{\partial V}\leq 0$, the sign of $\Delta G$ coincides with the sign of $\delta\mu$.

Let us now consider the effect of this correction on the STS results on some examples. 

\section{Examples}
We use the WKB approximation for the trapezoidal barrier  on Fig. \ref{model} for our calculations:
\begin{equation}
T=\exp\left(-z\frac{2\sqrt{2m}}{\hbar}\sqrt{\phi + \frac{V}{2}-E}\right)
\label{eq:T}
\end{equation}
and
\begin{equation}
\frac{\partial T}{\partial V}= - z\frac{\sqrt{2m}} {2\hbar \sqrt{\phi + \frac{V}{2}-E}}T,
\label{eq:deriv}
\end{equation}
where $\phi=\frac{\phi_s+\phi_t}{2}$ is the average work function of the tip and the sample. 
Substituting  (\ref{eq:deriv}) into (\ref{eq:deltaG}) and assuming that $V\ll \phi$ and the barrier transmission varies weakly on the scale of the gap in the density of states, we obtain an approximate expression for $d\tilde I/dV$
\begin{equation}
\frac{d\tilde I}{dV}\approx \left(\rho_s(V)+ z \sqrt{\frac{m} {2\hbar^2 \phi}} \int_0^\mu \rho_s(E)dE \right)AT,
\label{eq:vsz}
\end{equation}
which may be used to evaluate the correction.

\subsection{V-shaped LDOS}
\begin{figure}
\includegraphics[width=0.23\textwidth]{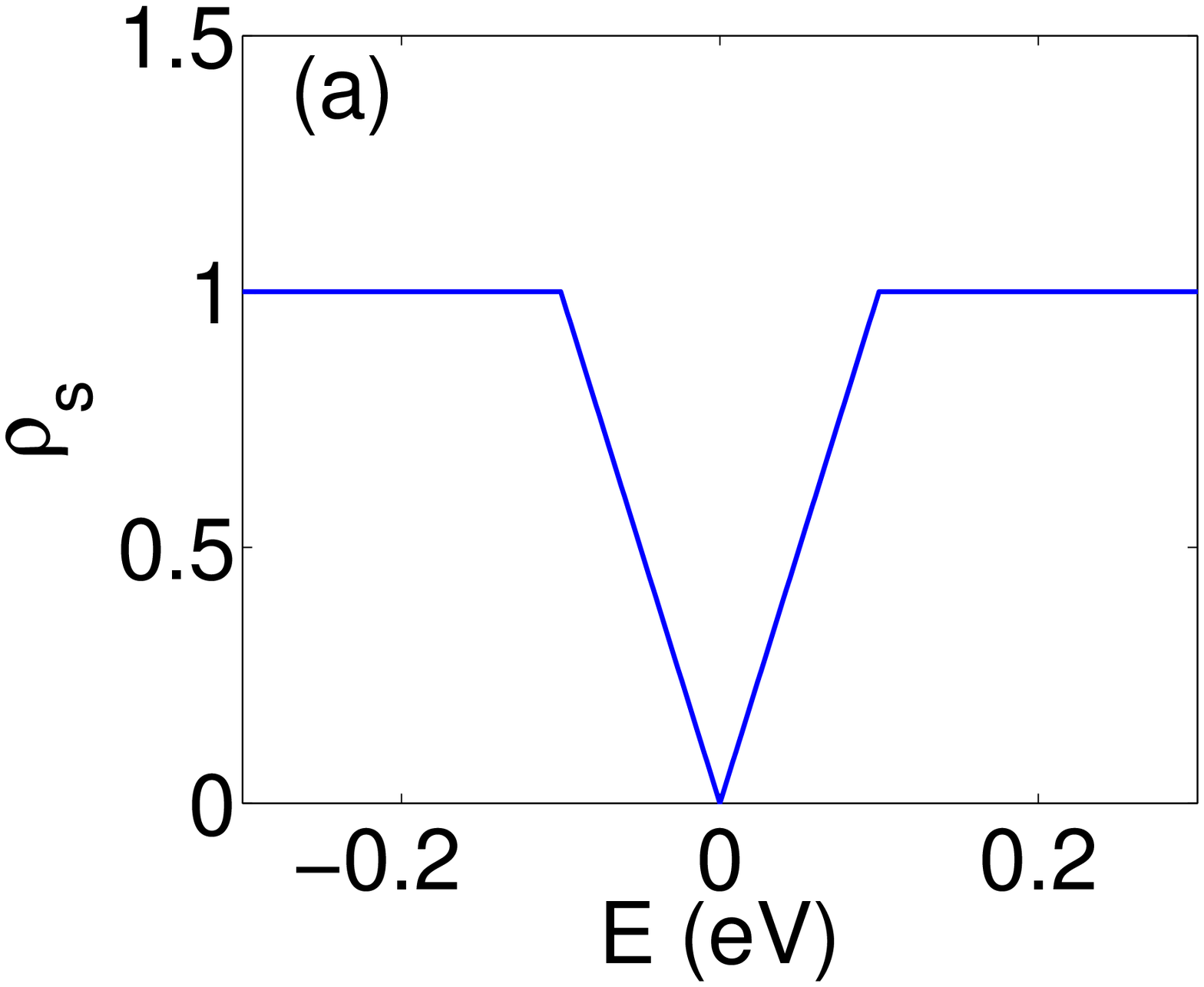}
\includegraphics[width=0.23\textwidth]{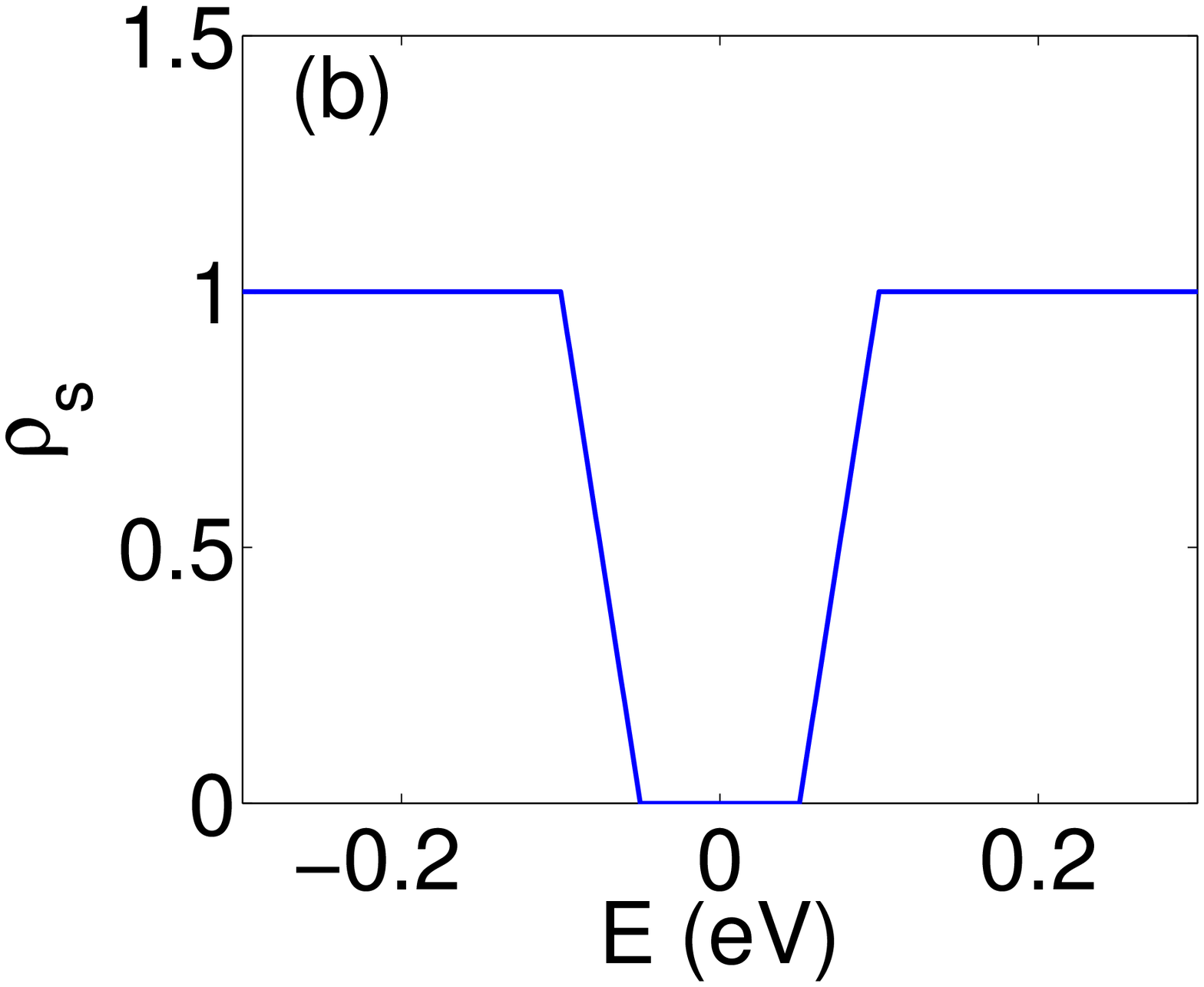}
\caption{Model LDOS: a) with a V-shaped feature with a width of 0.2 eV and  b) with an energy gap of 0.1 eV.}
\label{ldos1}
\end{figure}

We begin our analysis with a V-shaped density of states, shown in Fig.~\ref{ldos1}(а). 
Such a behavior exhibits a LDOS in the vicinity of the Dirac point of the surface states in 3D topological insulators. 
We still assume the temperature to be 0, tip density of states to be constant, and  we measure the value of the chemical potential from the minimum of the LDOS (\emph{i.e.} from the Dirac point in the case of 3D topological insulators). 

\begin{figure}
\includegraphics[width=0.45\textwidth]{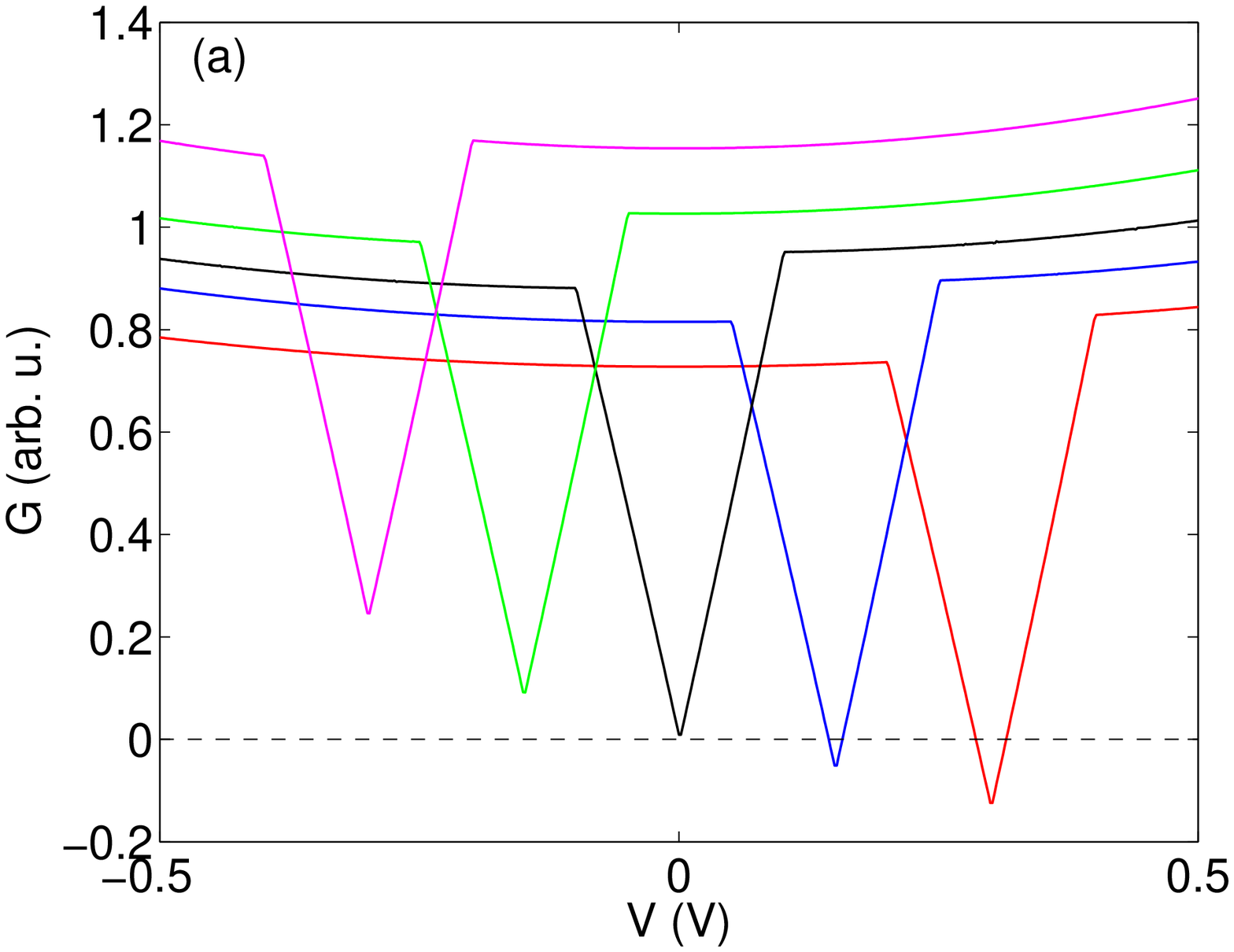}
\includegraphics[width=0.45\textwidth]{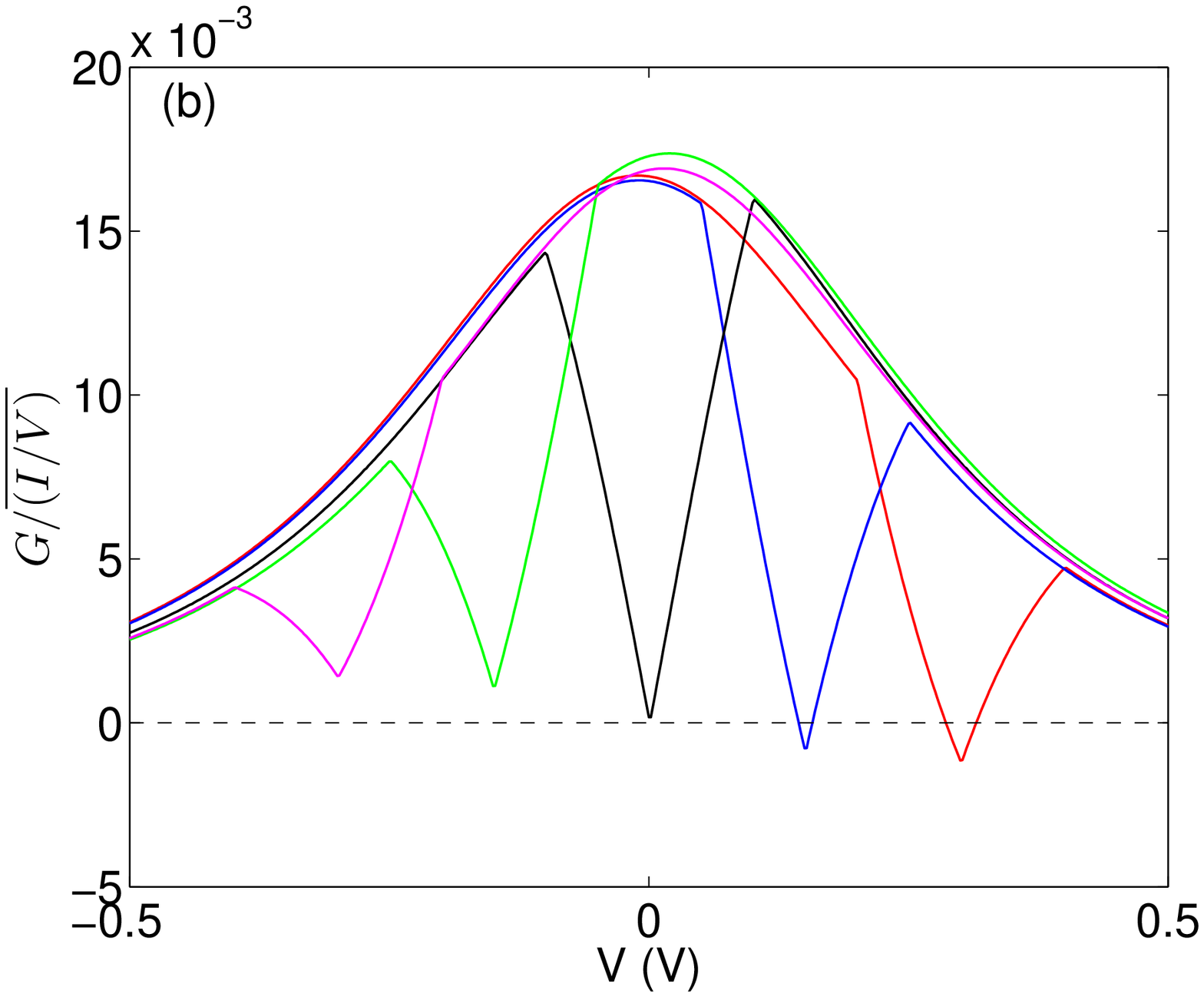}
\caption{$dI/dV$ и $d\ln I/d\ln V$ for the LDOS in Fig.~\ref{ldos1}(a), at $\delta\mu =-0.3$, -0.15, 0, 0.15 and 0.3~eV. $\phi_s =3.0$~eV, $\phi_t=5.0$~eV, $ z = 1$~nm.}
\label{vah1}
\end{figure}

In  Fig.~\ref{vah1}  a set of  $\frac{dI}{dV}(V)$ curves is shown modeled under these assumptions at different $\delta\mu$.
$\frac{dI}{dV}(V=0)|_{\delta \mu =0}=0$ in agreement with the usual interpretation of STS results.
However when the chemical potential shifts into the conduction band the value of the differential tunneling conductance at the minimum of the density of states grows and is not zero anymore, while when the chemical potential shifts into the valence band we observe negative differential conductance at the minimum of the sample LDOS, which in scanning tunneling spectroscopy is often considered a sign of poor quality of the tunneling probe. We also note, that the effect is significant and the apparent density of states may constitute a sizable fraction of the LDOS outside the gap. Obviously normalization by  $I/V$ does not lead to the disappearance of the effect, as is evident from Fig. \ref{vah1}(b). 

\begin{figure}
\includegraphics[width=0.45\textwidth]{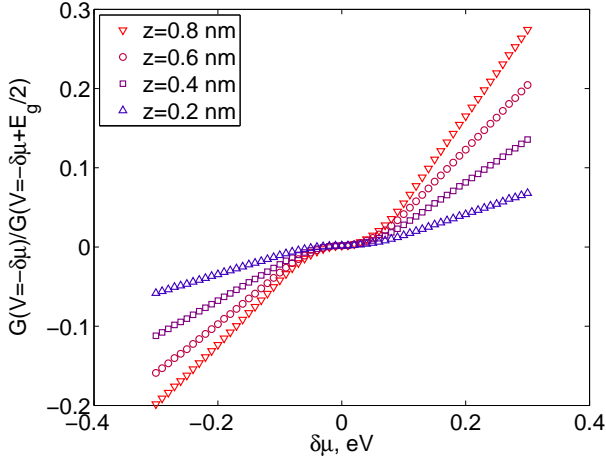}
\caption{The value of $dI/dV$ at the LDOS minimum (Fig.~\ref{ldos1}(a)), normalized by  $dI/dV$ at the gap edge, at different tip-sample separations.}
\label{gd1}
\end{figure}

 The dependence of $\frac{dI}{dV}(\delta \mu)$ at the minimum of the LDOS ($eV=-\delta \mu$) on  the chemical potential shift $\delta\mu$ is shown in Fig.~\ref{gd1}. It is evident that this dependence is nonlinear, has a slope reduction in the region of reduced density of states, and outside the V-shaped gap the dependence is stronger.

\subsection{Gapped LDOS}
\begin{figure}
\includegraphics[width=0.45\textwidth]{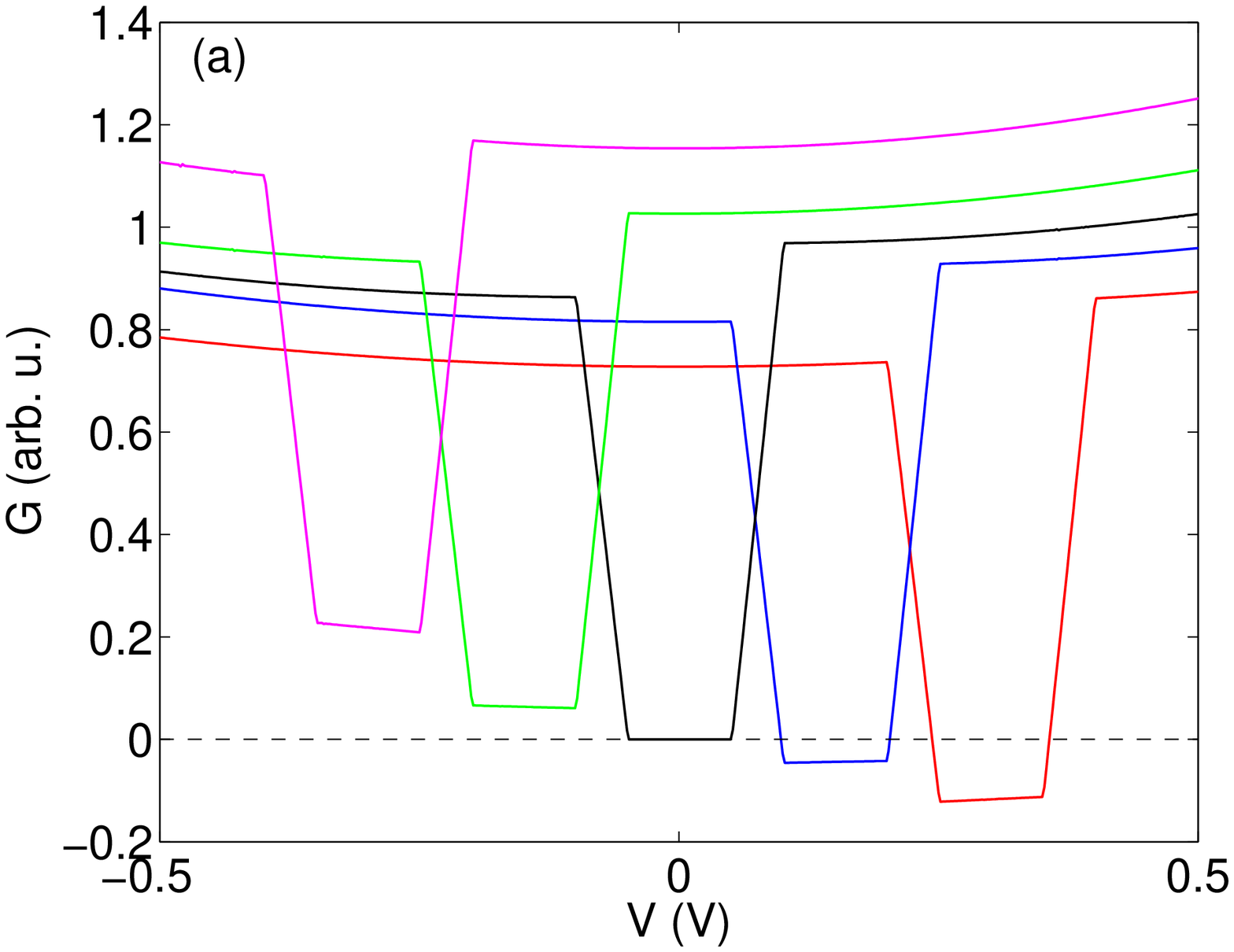}
\includegraphics[width=0.43\textwidth]{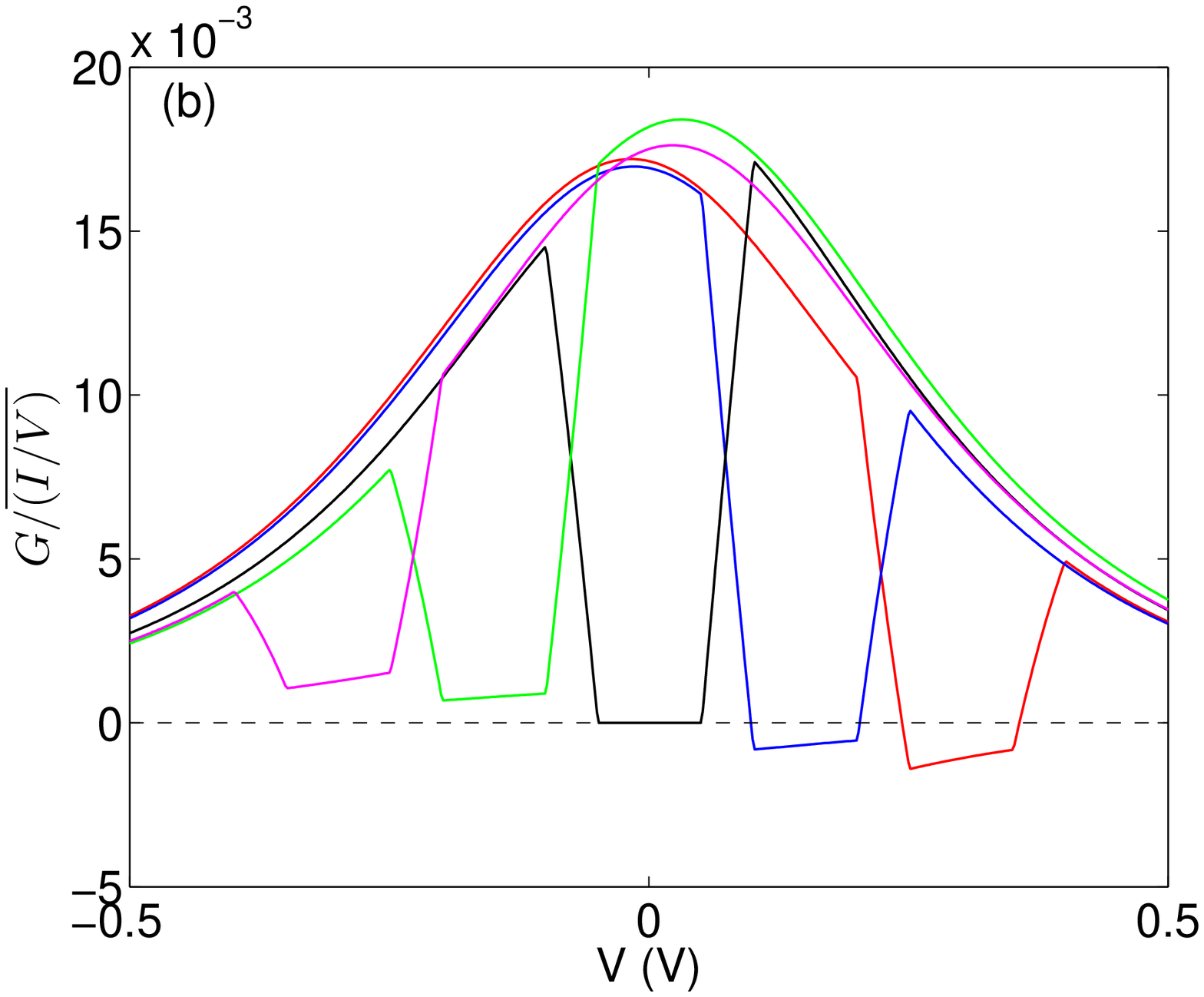}
\caption{$dI/dV$ (a) and $d\ln I/d\ln V$ (b) for the density of states in fig.~\ref{ldos1}(b),  at $\delta\mu =-0.3$, -0.15, 0, 0.15 and 0.3~eV. $\phi_s = 3.0 $~eV, $ \phi_t=5.0 $~eV, $ z = 0.6$~nm.}
\label{vah3}
\end{figure}

Let a gap opens in our V-shaped spectrum as in Fig. \ref{ldos1}(b). 
Such a density of states corresponds, for example, to a dielectric surface of a semiconductor, or a surface of a thin layer of a topological insulator where topologically protected surface states of opposite surfaces hybridize. 
A set of tunneling spectra obtained for such a LDOS is shown in Fig. \ref{vah3}.
 One can see, that zero density of states in the gap corresponds to nonzero differential tunneling conductance: positive if the chemical potential is in the conduction band (which might be incorrectly  interpreted as appearance of states inside the energy gap) and negative if it lies in the valence band (which might be incorrectly interpreted as a bad tunneling probe).  Negative differential conductance was observed in tunneling experiments on thin organic films and was attributed among other reasons to the change in the barrier transmission \cite{Wagner}. 
 
 As in the case of V-shaped density of states, the normalization of the tunneling spectra   $dI/dV$ by $I/V$ (Fig.~\ref{vah3}(b)) does not eliminate the correction. We also note a small tilt in the gap region on the    $\frac{dI}{dV}(V)$ and $d\ln I/d\ln V$ curves if the chemical potential is outside the gap, the sign of the tilt being opposite for $\frac{dI}{dV}(V)$ and $d\ln I/d\ln V$. The presence of such a tilt can be regarded as an indicator of a nonzero  $\Delta G$.
As expected, zero density of states in the  gap region corresponds to zero differential tunneling conductance only if the Fermi level is inside the gap. 

\begin{figure}
\includegraphics[width=0.45\textwidth]{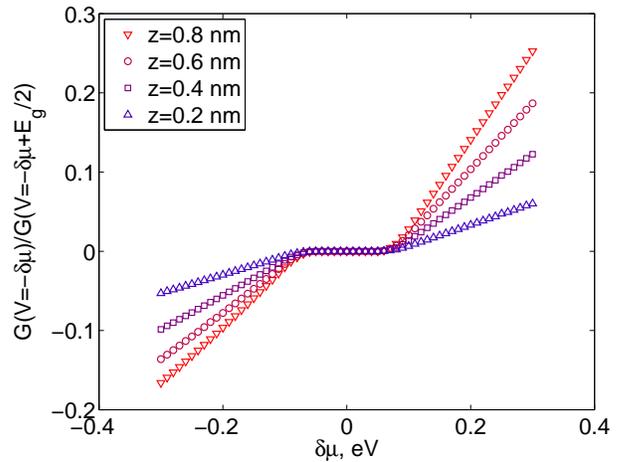}
\caption{The value of $dI/dV$ in the middle of the gap (Fig.~\ref{ldos1}(b)), normalized by  $dI/dV$ at the gap edge, at different tip-sample separations.}
\label{gd3}
\end{figure}

\subsection{Dependence of $\Delta G$ on $z$}
\begin{figure}
\includegraphics[width=0.4\textwidth]{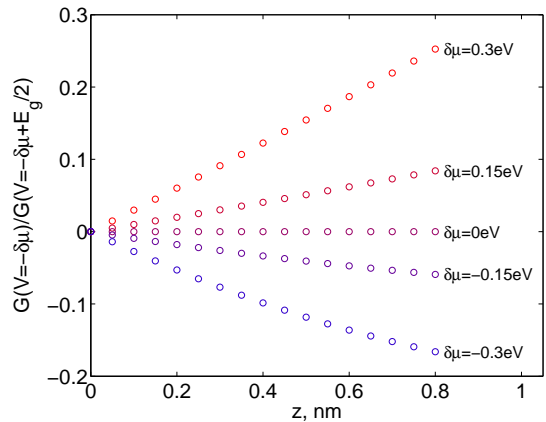}
\caption{The value of $dI/dV$ in the middle of the gap (Fig.~\ref{ldos1}(b)), normalized by  $dI/dV$ at the gap edge {\em vs.} tip-sample separation at different  positions of the chemical potential level.}
\label{gz3a}
\end{figure}

As it is evident from Figs.~\ref{gd1}, \ref{gd3}, the relative value of the correction depends on the distance between the probe and the sample.
A set of such dependences at different chemical potential positions is shown in Fig.~\ref{gz3a}.  In accordance with equation (\ref{eq:deriv}) the correction to $dI/dV$ diminishes almost linearly with decrease of the distance between the tip and the sample surface, with  $\lim_{z\to 0}\Delta G=0$. This property can be used to identify the presence of the contribution (\ref{eq:deltaG}), and also for its minimization or practically full elimination. 

Note, that almost linear dependence of the relative differential conductivity correction on a barrier width is not a result of the particular barrier shape (\ref{eq:T}). Since in practice the dependence of a tunneling current on a tip-sample separation is close to the exponential one \cite{Feenstra,Bai,Chen,STS,Voigtlaender}, the barrier transmission can be written as $T\propto \exp (-z f(E,V,z))$, where the dependence of $ f(E,V,z)$ on $z$ is weak. In this case $\partial T/\partial V\propto -z(\partial f/\partial V) T$, \emph{i.e.} the dependence of the correction on  $z$ should indeed be close to a linear one.

\section{Conclusion}
Thus, the dependence of the tunneling barrier transmission on applied voltage can result in  an additional contribution to the differential conductivity of the tunneling junction in the energy gap region. 
If the gap edge is reached with a nonzero value of a tunneling current, a change in the tunneling barrier transmission with voltage leads to a change of the tunneling current and a nonzero slope $dI/dV$ of the tunneling $I(V)$ curve, which in the usual approach is interpreted as nonzero density of states.
 In particular, this may explain substantially different values of normalized differential conductivity in the Dirac point of the tunneling spectra of the surface states of Bi$_2$Se$_3$ topological insulator (see, \emph{e.g.},  \cite{Dai}). Another example is investigation of surface states of semiconductors by low temperature scanning tunneling microscopy and spectroscopy. Common practice in this case is the usage of heavy doped degenerate semiconductors to exclude freezing out of current carriers.  This is a kind of systems where we can expect a significant contribution if the chemical potential position stays outside the energy gap in the surface density of states. To identify this contribution and eliminate it one can use almost linear dependence  of its normalized value on the tip-sample separation. On the other hand, if the gap opens on the Fermi level, as it occurs in superconductors, quasi-one dimensional conductors with charge or spin density waves and systems with Coulomb blockade, such a correction of zero value of the differential tunneling conductance does not emerge. 

This work was supported by RFBR (project 16-02-0067716), and programs of Physical Science Department of RAS and Presidium of RAS.

\end{document}